\documentclass[12pt,twoside]{article}
\usepackage{fancyheadings}
\usepackage{epsfig}
\usepackage{color}
\usepackage{wtopi}

\newcommand{\del}[1]{\delta\left(#1\right)}
\newcommand{\lap}{\nabla^2}

\newcommand{\rvec}[1]{\mathbf{#1}}

\newcommand{\ep}{\varepsilon}

\newcommand{\dprime}{{\prime\prime}}

\renewcommand{\Pi}{\mathbf{P_1}}

\newcommand{\x}{\mathbf{x}}
\newcommand{\xo}{\mathbf{x_0}}
\newcommand{\Gt}{\tilde{G}}

\newcommand{\fint}{\int_{-\infty}^{\infty}}
\newcommand{\hint}{\int_0^{\infty}}

\newcommand{\zp}{z^\prime}

\newcommand{\zo}{z_0}
\newcommand{\w}{\omega}

\newcommand{\xp}{x^\prime}
\newcommand{\yp}{y^\prime}
\newcommand{\xr}{x_R}
\newcommand{\xs}{x_S}
\newcommand{\Dt}{\tilde{D}}
\newcommand{\Vt}{\tilde{V}}
\newcommand{\kx}{k_x}
\newcommand{\ky}{k_y}
\newcommand{\kz}{k_z}
\newcommand{\kyp}{k_y^\prime}

\newcommand{\kxp}{k_x^\prime}

\newcommand{\km}{k_m}
\newcommand{\kh}{k_h}
\newcommand{\kp}{k^\prime}
\newcommand{\kmp}{k_m^\prime}
\newcommand{\khp}{k_h^\prime}
\newcommand{\khpp}{k_h^\dprime}
\newcommand{\sinc}{\mathrm{sinc}}
\newcommand{\Real}{\mathrm{Re}}
\newcommand{\Imag}{\mathrm{Im}}

\newcommand{\beq}{\begin{equation}}
\newcommand{\eeq}{\end{equation}}
\newcommand{\bea}{\begin{eqnarray}}
\newcommand{\eea}{\end{eqnarray}}

\begin{document}
\pagestyle{fancyplain}
\lhead[\thepage\hspace{10mm}\sl R.~B.~Schlottmann]{}
\rhead[]{{\sl Extended diffraction tomography}\hspace{10mm}\thepage}
\cfoot{}

\title{Extended diffraction tomography}

\author{R.~Brian Schlottmann}

\maketitle

%%%%%%%%%%%%%%%%%%%%%%%%%%%%%%%%%%%%%%%%%%%%%%%%%%%%%%%%%%%%%%%%%%%%%%%%%%%%%%%%%%%%%%%%%%%%%%%%%%%%%%%%%
\section*{Summary}

We present the development of extended diffraction tomography, a new approach to the solution of the linear seismic
waveform inversion problem.  This method has several appealing features, such as the use of arbitrary depth-dependent
reference models and the decomposition of the full 2D or 3D inverse problem into a large number of independent 
1D problems.  This decomposition makes the method naturally highly parallelizable.  Careful implementation 
yields significant robustness with respect to noise.  Several synthetic examples are shown which characterize the
benefits of our method and demonstrate the usefulness of choosing realistic 1D reference media.

%%%%%%%%%%%%%%%%%%%%%%%%%%%%%%%%%%%%%%%%%%%%%%%%%%%%%%%%%%%%%%%%%%%%%%%%%%%%%%%%%%%%%%%%%%%%%%%%%%%%%%%%%
\section{Introduction}

There is a vast amount of information contained in seismic data about the Earth structure through which the waves have traveled.
Traditional data processing has focused on the imaging of discontinuities---migration---and the reconstruction of the smooth 
component of the wave velocity variations through the tomographic analysis of traveltimes.  However, the use of the complete seismic
record to obtain high resolution velocity models through full-waveform inversion has long been limited by both the computational
cost and by a lack of theoretical guidance on this nonlinear problem.  In recent years, the rapid increase in computational power 
has enabled the development and implementation of some advances in this area, e.g. gradient techniques like those developed by 
Pratt (1999a, 1999b) and Sirgue and Pratt (2004).  However, these methods are slow to converge and hard to implement in three dimensions. 

More recently, advances in the area of inverse scattering theory (Weglein {\em et al.}, 2003; Schlott\-mann, 2006) have arisen that 
may have the potential of solving the nonlinear full-waveform inversion problem more quickly and efficiently.  However, these methods
rely on having an efficient and reliable method of solving the {\em linear} waveform inversion problem, i.e., the problem that results
from assuming that the wave propagation can be approximated by first-order scattering, yielding a linear relationship between the 
data and the unknown subsurface structure.

The broad class of techniques that address this linear problem fall under the rubric of ``diffraction tomography''.  Although the
literature on this subject is vast, our work is chiefly influenced by Wu and Toks\H{o}z (1987).  In that work the authors extend
earlier work of Devaney (1984) and Devaney and Oristaglio (1984) to linearly invert surface reflection profile (SRP) data assuming
point sources.  Wu and Toks\H{o}z essentially treat the linear relationship that maps subsurface structure into data as an integral 
transform and proceed to develop a partial inverse transform.  In this paper, we will refer to their result as ``classical diffraction 
tomography''.

The inverse transform of classical diffraction tomography is partial in the sense that not all of the subsurface structure can be 
recovered under realistic conditions.  Specifically, in the absence of very low-frequency data---the norm in SRP data---many long 
wavelength components of the structure are irretrievable.  The origin of this problem is closely tied to the assumption that 
the true subsurface is close to a homogeneous medium and only differs from such by a small enough degree that the linear first-order
scattering approximation is valid.  This assumption of a homogeneous background or reference medium in addition to the lack of 
resolution of long wavelengths greatly limits the applicability of this method.

However, classical diffraction tomography does have the advantage of being an efficient and stable method.  It is desirable to preserve
those qualities while seeking a new theory that frees us from the assumption of a homogeneous reference and enables significantly
better resolvability of subsurface structure.

In this paper, we present the development of such a result.  Our method allows for any 1D (i.e. depth-dependent) reference medium
while also decoupling the full 2D or 3D inverse problem into a large number of independent 1D inverse problems, resulting in 
an efficient algorithm that is naturally highly parallelizable.  With careful implementation, the algorithm that is obtained from our
theory is stable with respect to noise.  By assuming even a marginally realistic 1D reference medium, the ability to resolve 
long wavelength components of the subsurface without unrealistically low frequencies is greatly improved.

%%%%%%%%%%%%%%%%%%%%%%%%%%%%%%%%%%%%%%%%%%%%%%%%%%%%%%%%%%%%%%%%%%%%%%%%%%%%%%%%%%%%%%%%%%%%%%%%%%%%%%%%%
\section{Theory}

\subsection{Two dimensions}

For the moment, we limit ourselves to a seismic surface reflection profile in only two dimensions.  We begin by assuming 
that we measure the scattered 
component of the seismic wavefield over a finite region of the surface.  We also assume that the data have been organized in the
midpoint-offset domain.  Specifically, if the data were recorded at horizontal shot/receiver locations $\xs,\xr$ (all located at 
depth $z=0$), we order the data in terms of the coordinates $m,h$:
\bea
m&=&\frac{1}{2}(\xr+\xs)\nonumber\\
h&=&\frac{1}{2}(\xr-\xs),
\eea
subject to the limitations
\bea
|m|&\leq & a\nonumber\\
|h|&\leq & b.
\eea
In the case of most seismic surveys, such as a standard marine survey, the range of the midpoint $m$ is controlled by the 
length of the survey line, whereas the range of $h$ is controlled by the aperture of the survey, e.g. the length of a streamer
cable.  

Since our objective is to develop an accurate, efficient, and stable solution of the {\em linear} inverse problem, we take the scattered 
component of our wavefield to be due to first-order Born scattering.  In the frequency domain, assuming constant density and acoustic
propagation, we have
\beq
D(\xr,\xs)=\w^2\fint dx\hint dz\ G(\xr,z_R=0,\w;x,z)V(x,z)G(x,z,\w;\xs,z_S=0)\label{Born}
\eeq
where the scattering potential $V$ is defined to be 
\beq
V(x,z)=\frac{1}{c^2(x,z)}-\frac{1}{c_0^2(z)}.
\eeq
The wave velocities $c$ and $c_0$ are those of the actual and reference media, respectively, and the reference medium is
assumed to vary only with depth.  The function $G(\x,\w;\xo)$ is the Green's function of the reference medium and is the solution to 
\beq
-\frac{\w^2}{c_0^2(z)}G=\lap G+\del{\x-\xo}.
\eeq
Because our reference medium is laterally invariant, $G$ has the important and useful property
\beq
G(x,z,\w;x_0,\zo)=G(|x-x_0|,z,\w;0,\zo).\label{Invariant}
\eeq
Note that the inclusion of a {\em flat} free surface will not destroy this property.  Using this invariance, we rewrite eq.~(\ref{Born})
in the $m,h$ system (with the source/receiver depths suppressed):
\beq
D(m,h)=\w^2\fint dx\hint dz\ G(m+h-x,z,\w)V(x,z)G(m-h-x,z,\w).\label{BornMH}
\eeq
This is the form with which we will work.

If the wavefield is sampled densely enough relative to the frequency range we consider, we can treat $m$ and $h$ as continuous 
variables.  Doing so, we take the Fourier transform of the data over the given ranges of $m$ and $h$:
\beq
\Dt(\km,\kh)=\frac{1}{2\pi}\int_{-a}^{a} dm \int_{-b}^{b} dh\ e^{-i\km m} e^{-i\kh h} D(m,h).
\eeq
(Note that we use a symmetric normalization of $1/\sqrt{2\pi}$ in our forward and inverse Fourier transforms.)  We also take
\beq
G(x-x_0,z,\w;0,0)=\frac{1}{\sqrt{2\pi}}\fint dk\ e^{ik(x-x_0)} \Gt(k,z,\w).
\eeq
Using both of the above in eq.~(\ref{Born}) and performing the integrals over $m$ and $h$, we get
\bea
\Dt(\km,\kh)&=&\frac{\w^2 a b}{\pi^2}\fint dx \hint dz \fint dk \fint d\kp\ 
\sinc\left[(\km-k-\kp)a\right] \sinc\left[(\kh-k+\kp)b\right]\nonumber\\
& &\hspace{65mm}\times e^{-ix(k+\kp)}V(x,z)\Gt(k,z,\w)\Gt(\kp,z,\w)\label{Dkmh1}
\eea
Making the change-of-variables
\bea
\kmp&=&k+\kp,\nonumber\\
\khp&=&k-\kp,\label{COV}
\eea
(which has a Jacobian determinant of -$\frac{1}{2}$) and performing the integral over $x$, we get
\bea
\Dt(\km,\kh)&=&\frac{\w^2 a b}{2\pi^2}\sqrt{2\pi}\hint dz \fint d\kmp \fint d\khp\ 
\sinc\left[(\km-\kmp)a\right] \sinc\left[(\kh-\khp)b\right]\Vt(\kmp,z)\nonumber\\
& &\hspace{50mm}\times \Gt\left[\frac{1}{2}(\kmp+\khp),z,\w\right]\Gt\left[\frac{1}{2}(\kmp-\khp),z,\w\right].\label{Dkmh}
\eea

For our purposes, the first sinc function in eq.~(\ref{Dkmh}) is very inconvenient because it couples together all the horizontal
wavenumbers of the unknown potential $\Vt$, breaking what would otherwise be a complete decoupling of the inverse problem for each
value of $\km$.  Fortunately, seismic surveys tend to be done over fairly large horizontal areas, which means that $a$ can be
expected to be large (unlike the aperture).  As $a$ tends to infinity, the first sinc function will tend towards a $\delta$-function.
However, in many cases the survey length will still not be large enough to allow the $\delta$-function approximation, but we can 
make another modification to rectify the situation.  Specifically, by adding a small imaginary part to the frequency,
\beq
\w\rightarrow\w+i\ep,
\eeq
we can force the wavefield to decay rapidly enough that we can legitimately make the substitution
\beq
a\sinc\left[(\km-\kmp)a\right]\rightarrow\pi\del{\km-\kmp},
\eeq
in effect treating our survey length as infinite.  Of course, the addition of the imaginary part to the frequency is equivalent to
tapering our data in the time domain with a slowly decaying exponential.  In practice, the imaginary part will not need to be much more 
than a few percent of the real part of the frequency.

What we have gained for our efforts so far is that we can now write the relationship between our data and the scattering potential as
\beq
\Dt(\km,\kh)=\hint dz\ M(\kh,z;\km)\Vt(\km,z)\label{DMV}
\eeq
where
\beq
M(\kh,z;\km)=\frac{\w^2b}{\sqrt{2\pi}}\fint d\khp\ \sinc\left[(\kh-\khp)b\right]
\Gt\left[\frac{1}{2}(\km+\khp),z,\w\right]\Gt\left[\frac{1}{2}(\km-\khp),z,\w\right].\label{M}
\eeq
Thus, what we have achieved is the transformation of the 2D inverse problem into a set of independent 1D inverse problems, one for 
each horizontal wavenumber $\km$.   

Before we can use our results, we must consider an additional issue.  The symmetry of the reference Green's function $G$ with
respect to the horizontal offset $x-x_0$, i.e., its dependence on the absolute value $|x-x_0|$, yields
the same symmetry in the horizontal wavenumber domain:
\beq
\Gt(k,z,\w)=\Gt(-k,z,\w).
\eeq  
This property, after a little thought, implies that $M$ in eq.~(\ref{M}) is symmetric in $\km$:
\beq
M(\kh,z;\km)=M(\kh,z;-\km).
\eeq
Because of this and because the data have no particular symmetry with respect to $\km$, if we were simply to use eq.~(\ref{DMV}), 
the appropriate symmetry of the function $\Vt$, which is the Fourier transform
of a real-valued function, would not be preserved.  In other words, we must reorganize the problem to guarantee that
\beq
\Vt(\km,z)=\Vt^{*}(-\km,z).\label{VSym}
\eeq

To do this, we take eq.~(\ref{DMV}) and itself evaluated at $-\km$ and either add or subtract them to get two equations:
\bea
\hint dz\ M(\kh,z;\km)\Real\{\Vt(\km,z)\}&=&\frac{1}{2}(\Dt(\km,\kh)+\Dt(-\km,\kh))\nonumber\\
\hint dz\ M(\kh,z;\km)\Imag\{\Vt(\km,z)\}&=&\frac{1}{2i}(\Dt(\km,\kh)-\Dt(-\km,\kh)).\label{Split}
\eea
This set of equations can be safely used to solve for $\Vt$ for all nonnegative $\km$, using eq.~(\ref{VSym}) to retrieve
the results for negative wavenumbers.

\subsection{A practical 3D geometry}

Ideally, the extension of our results in 2D to three dimensions would assume that for each shot, the wavefield would be recorded
over a 2D grid on the surface.  However, it is not usually practical to cover a large 2D surface area with a dense set of receivers, so 
we consider instead a 3D survey comprised of many 2D surveys.  Specifically, we will assume that we have a large number of data sets
like those we considered in the 2D case, where the sources and receivers are aligned along the $x$-direction, but each set will be
for a different value of the $y$-coordinate with the limitation $|y|\leq c$.  Thus, the new version of eq.~(\ref{Born}) is 
\beq
D(\xr,\xs,y)=\w^2\fint d\xp \fint d\yp \hint d\zp\ G(\xr,y,0,\w;\xp,\yp,\zp)V(\xp,\yp,\zp)G(\xp,\yp,\zp,\w;\xs,0)\label{Born3D}.
\eeq
We continue to use a 1D reference model, $c_0(z)$, so $G$ now has the property
\beq
G(x,y,z,\w;x_0,y_0,\zo)=G(|x-x_0|,|y-y_0|,z,\w;0,0,\zo).\label{Invariant3D}
\eeq

We proceed as in the 2D case and organize each survey line into the $m,h$ domain.  We again take the Fourier transform,
\beq
\Dt(\km,\kh,\ky)=\frac{1}{(2\pi)^{\frac{3}{2}}}\int_{-a}^{a} dm \int_{-b}^{b} dh \int_{-c}^{c} dy\ e^{-i\km m}
e^{-i\kh h} e^{-i\ky y} D(m,h,y),
\eeq
and expand $G$,
\beq
G(x,y,z,\w)=\frac{1}{2\pi}\fint d\kxp \fint d\kyp e^{i\kxp x} e^{i\kyp y} \Gt(\kxp,\kyp,z,\w).
\eeq
We also repeat our trick of adding an imaginary part to the frequency so that we can essentially take both the survey lengths
in both horizontal directions to be infinite.  The result, after implementing a double set of change-of-variables of the form
in eq.~(\ref{COV}), is
\bea
\Dt(\km,\kh,\ky)&=&\w^2 b\sqrt{\frac{\pi}{2}}\hint d\zp \fint d\khp \fint d\khpp\ \sinc\left[(\kh-\khp)b\right]\Vt(\km,\ky,\zp)\nonumber\\
& & \hspace{40mm} \times \Gt\left[(\frac{1}{2}(\km+\kmp),\frac{1}{2}(\ky+\khpp),\zp)\right]\nonumber\\
& & \hspace{40mm} \times \Gt\left[(\frac{1}{2}(\km-\kmp),\frac{1}{2}(\ky-\khpp),\zp)\right].
\eea
As in the 2D case, we have achieved a decoupling of the inverse problem.  Here we have broken down the full 3D problem into a large set
of independent 1D problems, one for each pair of $\km,\ky$ values.  Also as with the 2D result, care must be taken to ensure that
we account for the symmetries in the problem, similar to the procedure used to obtain eqs.~(\ref{Split}).  

As a final thought on this 3D treatment, we suggest that improved results would almost certainly be obtained by performing an additional
set of 2D surveys oriented perpendicular to the first.  These surveys, aligned along the $y$-direction, would provide a second,
independent set of equations for $\Vt$, most likely improving the condition of the inverse problem.

\section{Numerical Examples}

\subsection{Acquisition parameters and methodology}

In this section, we present several simple numerical examples designed to explore the accuracy and stability of extended diffraction
tomography.  In all cases we limit ourselves to the 2D case.   

In each case, the synthetic data is generated via first-order Born scattering (eq.~\ref{BornMH}) directly in the $m,h$ domain for 
frequencies of 1-10 Hz at a spacing of 1 Hz.  For all our examples, we have used the values $a=20$ km and $b=3$ km, indicating a 
total survey length of 40 km and an aperture of 6 km, both of which are reasonable values for modern marine acquisition geometries.  
We have assumed that $m$ and $h$ are regularly spaced at 50 m intervals.  All sources and receivers are taken to lie at depth $z=0$ km.

As mentioned in the last section, we add a small imaginary component to the frequency.  In all cases, that component was .15 
$\mathrm{s}^{-1}$, which implies an exponential taper in the time domain which drops by a factor of $e^{-1}$ in a characteristic 
time of 6.67 s.  Note that this value was determined by trial-and-error to be approximately the smallest effective one.

Although extended diffraction tomography can incorporate a flat free surface in its reference model, we did not do so in these examples 
but will include that additional bit of realism in future work.  We used two simple reference models to demonstrate the advantage of
being able to have a non-constant reference.  We begin with a homogeneous medium of constant velocity $c_0=1500$ m/s, i.e., water.  We 
follow that with a two-layer medium which has water everywhere above depth $z=1$ km and $c_1=4000$ m/s below.

To obtain our inversions, we discretized eqs.~(\ref{Split}), using the values of $\km$ and $\kh$ that naturally result from
discretizing $m$ and $h$.  The depth coordinate was chosen to range from $z=800-4000$ m with a grid spacing of 50 m.  Note that the
omission of $z$ values less than 800 m represents {\em a priori} information about $V$.  Although the dependence on frequency was
suppressed in eqs.~(\ref{Split}), we did use multiple frequencies {\em simultaneously} in each inversion, a procedure which, not 
surprisingly, augments both stability and accuracy.  However, the exact subset used was not always the same, as will be discussed later.

In most of our examples, we also demonstrate the usefulness of filtering in the vertical wavenumber (or $\kz$) domain.  The idea is to
limit the inversion so that the 2D Fourier transform of $V$, $\Vt(\kx,\kz)$, is non-zero only within some disk around $\rvec{k}=0$. 
This is done by rewriting the integrals over $z$ in eqs.~(\ref{Split}) as integrals over $\kz$ and substituting $M$ and $\Vt$ with 
their Fourier transforms over $z$.
This is another imposition of an {\em a priori} bias on our solution, one which helps to augment the low-wavenumber part of the
solution, which tends to be the most difficult to reconstruct.  In our examples, the disk was limited to 
\beq
|\rvec{k}|\leq 2.2 \frac{\w_{max}}{c_0}
\eeq
where $\w_{max}$ was the largest angular frequency used in the inversion.  Of course, this choice was motivated by classical
diffraction tomography, where the largest value of $|\rvec{k}|$ obtainable is $2\w_{max}/c_0$, but in our case we expect to 
recover some of the ``energy'' in $\Vt(\kx,\kz)$ outside this envelope, so the limit was chosen to be 20\% larger.

Finally, we must mention that the numerical method we used to solve the linear eqs.~(\ref{Split}) was the well-known technique of
singular value decomposition, or SVD (see e.g. Press {\em et al.}, 1992).  SVD uses the fact that any $M\times N$ matrix $\rvec{A}$ such
that $M\geq N$ may be decomposed into the product 
\beq
\rvec{A}=\rvec{U}\ \rvec{S}\ \rvec{V^T}
\eeq
where $\rvec{U}$ is an $M\times N$ column-orthogonal matrix, $\rvec{V}$ is an $N\times N$ orthogonal matrix, and $\rvec{S}$ is an 
$N\times N$ diagonal matrix whose elements, known as the singular values, are all nonnegative.  With this decomposition, a solution
to the problem
\beq
\rvec{A}\rvec{x}=\rvec{d}
\eeq
may be obtained in a least squares sense:
\beq
\rvec{x}=\rvec{V}\ \rvec{S^{-1}}\ \rvec{U^{T}}\ \rvec{d}
\eeq

\noindent Since the singular values may be zero or very small, in practice one actually replaces $\rvec{S^{-1}}$ with another matrix 
$\rvec{\tilde{S}^{-1}}$ such that if $s_j$ is the $j$-th singular value, then
\beq
(\rvec{\tilde{S}^{-1}})_{jj}=\left\{\begin{array}{cc}
                 1/s_j,     &    s_j\geq\sigma\\
                 0,     &    s_j<\sigma
\end{array}\right.
\eeq
for some tolerance $\sigma$.
 
The selection of the tolerance is quite critical in our inversion problems.  Generally, the matrix equations that we must solve
are very poorly conditioned, with many singular values being very close to zero.  The main difficulty that arises from this poor
conditioning is an extreme sensitivity to noise.  This occurs because for very small $s_j$, the $j$-th diagonal element of
$\rvec{S}^{-1}$ is extremely large, amplifying any errors in our data, such as noise.  In our computations, a tolerance 
of $\sigma=1$ was used uniformly.  As we will see, this will provide robustness with respect to noise.  However, the price for this
robustness is some degradation in the quality of our inversion, as some features of $V$ which we would like to recover only weakly
affect the data through some of the small $s_j$ that we drop from consideration.  

\subsection{Examples}

\newlength{\wid}
\setlength{\wid}{4.0in}
\begin{figure}[!htbp]
\begin{center}
\epsfig{file=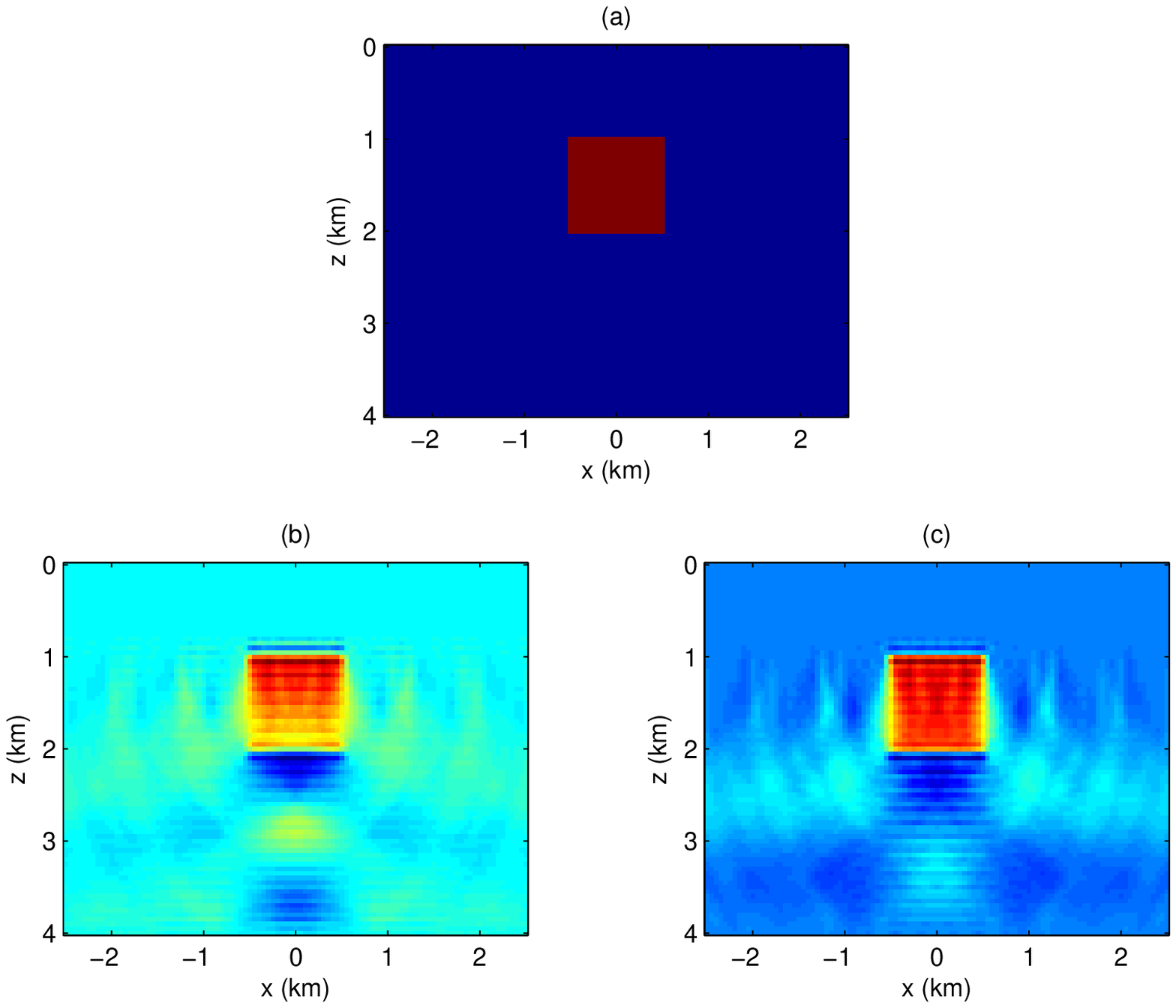,width=\wid,clip=}
\end{center}
\caption{Results for a homogeneous background using frequencies 1,2,...,5 Hz. (a) True model. (b) Result for unrestricted $k_z$.  (c)
Result for restricted $k_z$.}
\begin{center}
\epsfig{file=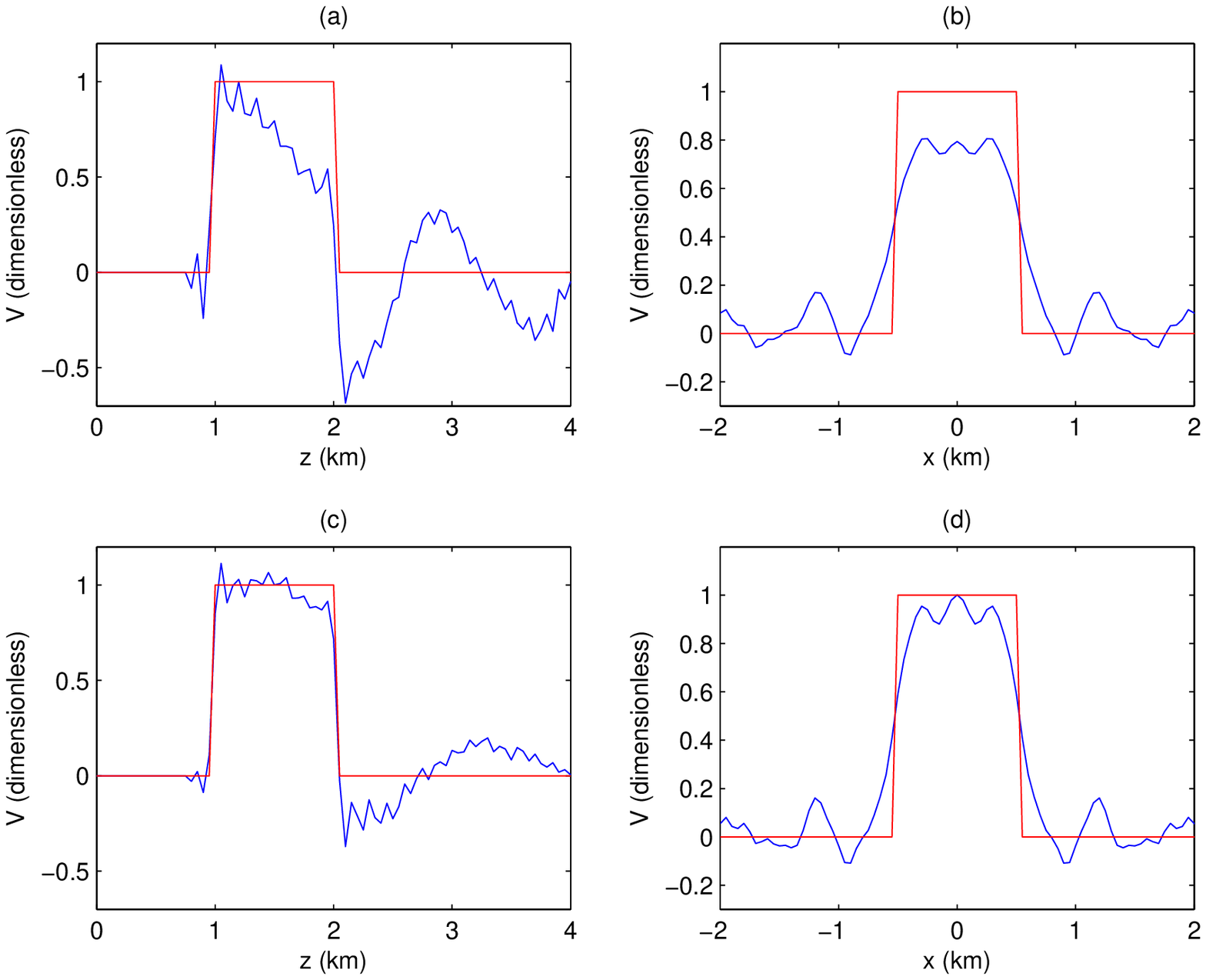,width=\wid,clip=}
\end{center}
\caption{Cross-sections through the centers of the squares in Fig. 1.  In each case, the true model is in red, and the
reconstruction is in blue.  (a,b) Vertical and horizontal slices through the unrestricted result.  (c,d) Vertical and horizontal
slices through the restricted results.}
\end{figure}

In all of the following figures, we will see images of both the true models used and the inversion results as well cross-sections 
through them.  In the model and inversion images, only the middle 4 km of the full 40 km $x$-coordinate range is shown in order to 
allow the inversion results to be more easily visible.  Although it is not indicated in the figures, as stated above, the sources and
receivers are assumed to be located at $z=0$, the top of the image boxes.  Each cross-section shown, vertical and horizontal, is taken 
through the center of the anomaly we are trying to reconstruct.   Note that in each such slice, the true model will always be 
plotted in red while the inversion results will be in blue.  

In all cases, the input model is a square-shaped anomaly of side length 1 km centered horizontally at $x=0$ but with different
depth for different examples.  The value of the scattering potential $V$, which is a dimensionless quantity, 
is always 1, corresponding to a low-velocity region.  However, since this is a purely linear inverse problem, the actual magnitude and
sign of $V$ are irrelevant.  This would most certainly not be the case for the nonlinear full-waveform problem.

We begin by looking at some results for a homogeneous reference medium.  In Figures 1 and 2, we see the results for our square 
centered at a depth of $z=1500$ m.  In this test, frequencies of 1-5 Hz were used.  Notice that there is marked improvement in the results
when the range of $\kz$ values is restricted.

In Figures 3 and 4 we have the results for the same homogeneous background but with the square now centered 500 m deeper at $z=2000$ m.  
Clearly, the inversion suffers serious degradation from even this slight increase in target depth.  We are already reaching the limit of 
our method for a constant velocity reference, even with $\kz$ restriction.  These images could be improved by lowering the tolerance 
$\sigma$, but then sensitivity to noise would grow rapidly.

\begin{figure}[!htbp]
\begin{center}
\epsfig{file=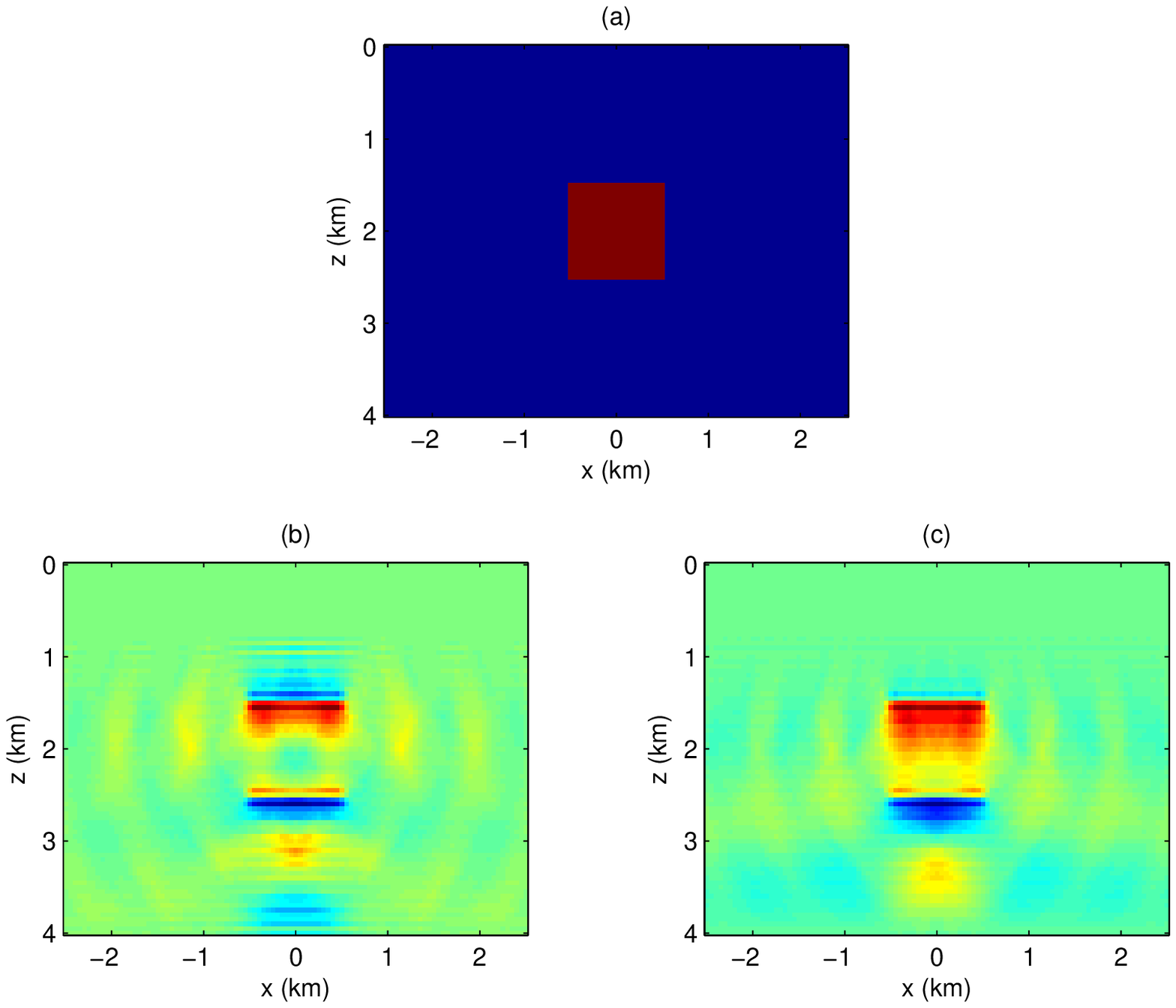,width=\wid,clip=}
\end{center}
\caption{Results for the same situation as in Fig. 1 except that the square is 500 m lower.  (a) True model.  
(b,c) Unrestricted and restricted
results.}
\begin{center}
\epsfig{file=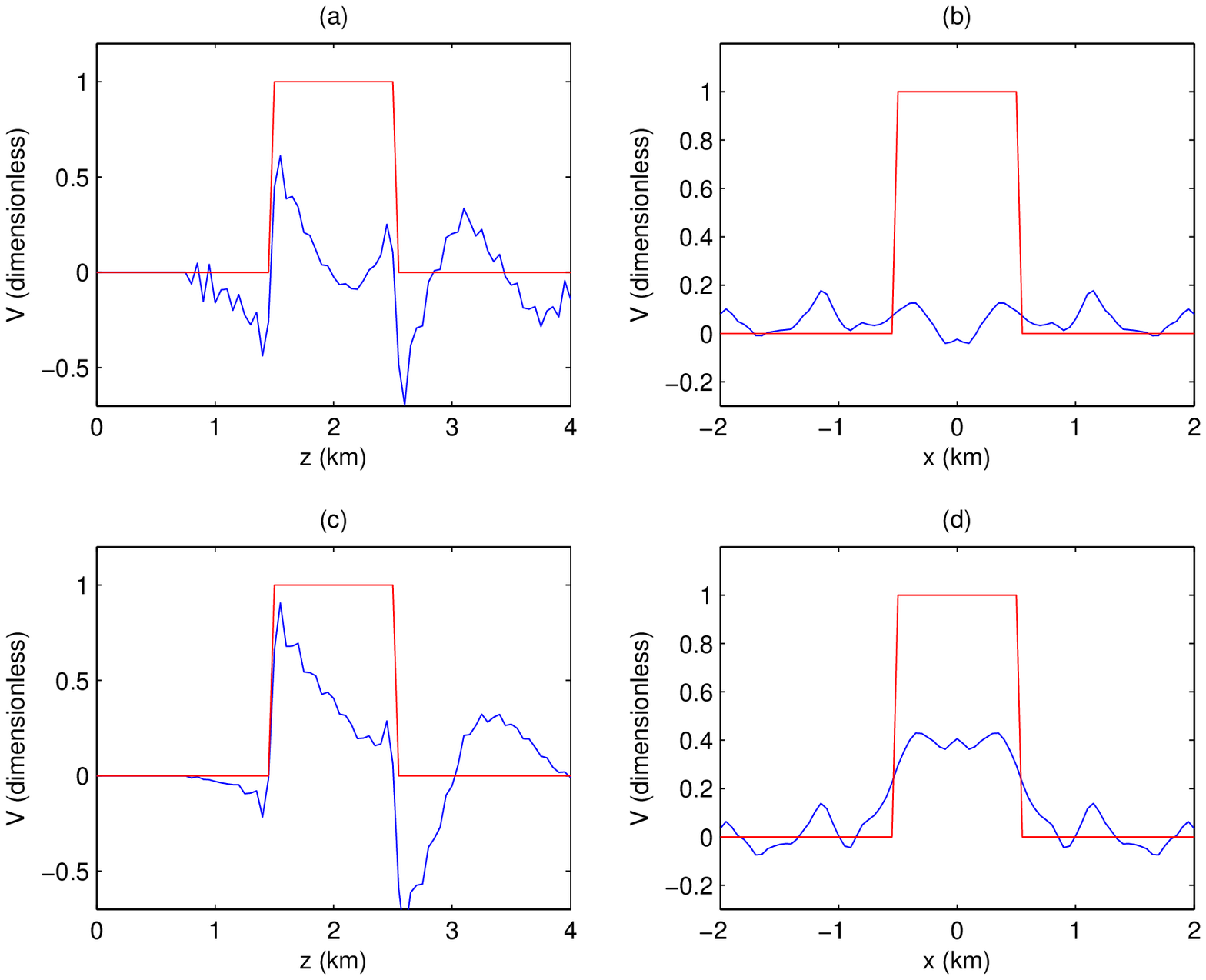,width=\wid,clip=}
\end{center}
\caption{Cross-sections through the results of Fig. 3.  (a,b) Vertical and horizontal slices through the unrestricted result.
(c,d) Vertical and horizontal slices through the restricted results.}
\end{figure}

Next, in Figures 5 and 6, we explore the effect of having a two-layer reference model on reconstructing the deeper anomaly.  
Note that Figure 5a shows the 
scattering potential $V$, while Figure 5b shows the complete velocity model.  With nothing else being changed from the previous
example, we see dramatic improvements in the image quality.  We attribute this improvement to two effects.  One is the ``stretching''
of the wavelengths of the incident waves as they enter the higher-velocity lower half-space.  Thus, for the same frequency you gain
sensitivity to lower-wavelength components of $V$.  (This actually results in slightly worse reconstruction of the higher wavelengths
of $V$.)  The other effect which adds to our image quality is the refraction of the incident waves, which brings them closer to 
horizontal, again improving our sensitivity to wavelength components that are hard to recover in a homogeneous reference model.
Note also that the restriction of $\kz$ adds little to the improvement of the results here.

In Figures 7 and 8, we see the advantage of adding higher frequencies into the inversion.  Here we have used the full range of
frequencies, 1-10 Hz, that were prepared for these tests.  Clearly, the anomaly is well-resolved at all wavelengths, and, again,
$\kz$ restriction adds little other than the dampening of small artifacts.

\begin{figure}[!htbp]
\begin{center}
\epsfig{file=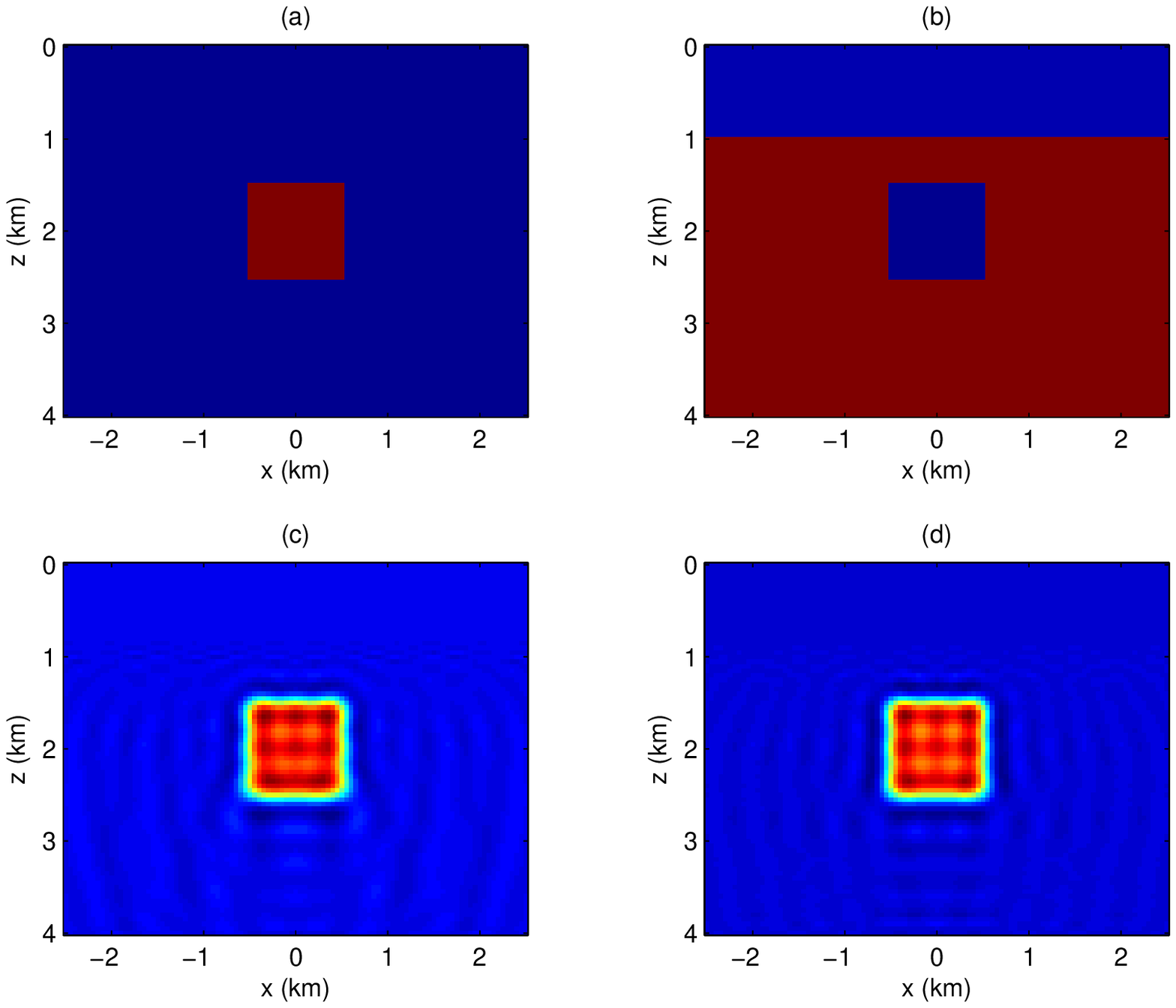,width=\wid,clip=}
\end{center}
\caption{Results for the two-layer background model. Again frequencies 1,2,...,5 Hz are used.  (a) True model of $V$.  (b) True velocity
model ($V$ plus background). (c,d) Results for unrestricted and restricted cases.}
\begin{center}
\epsfig{file=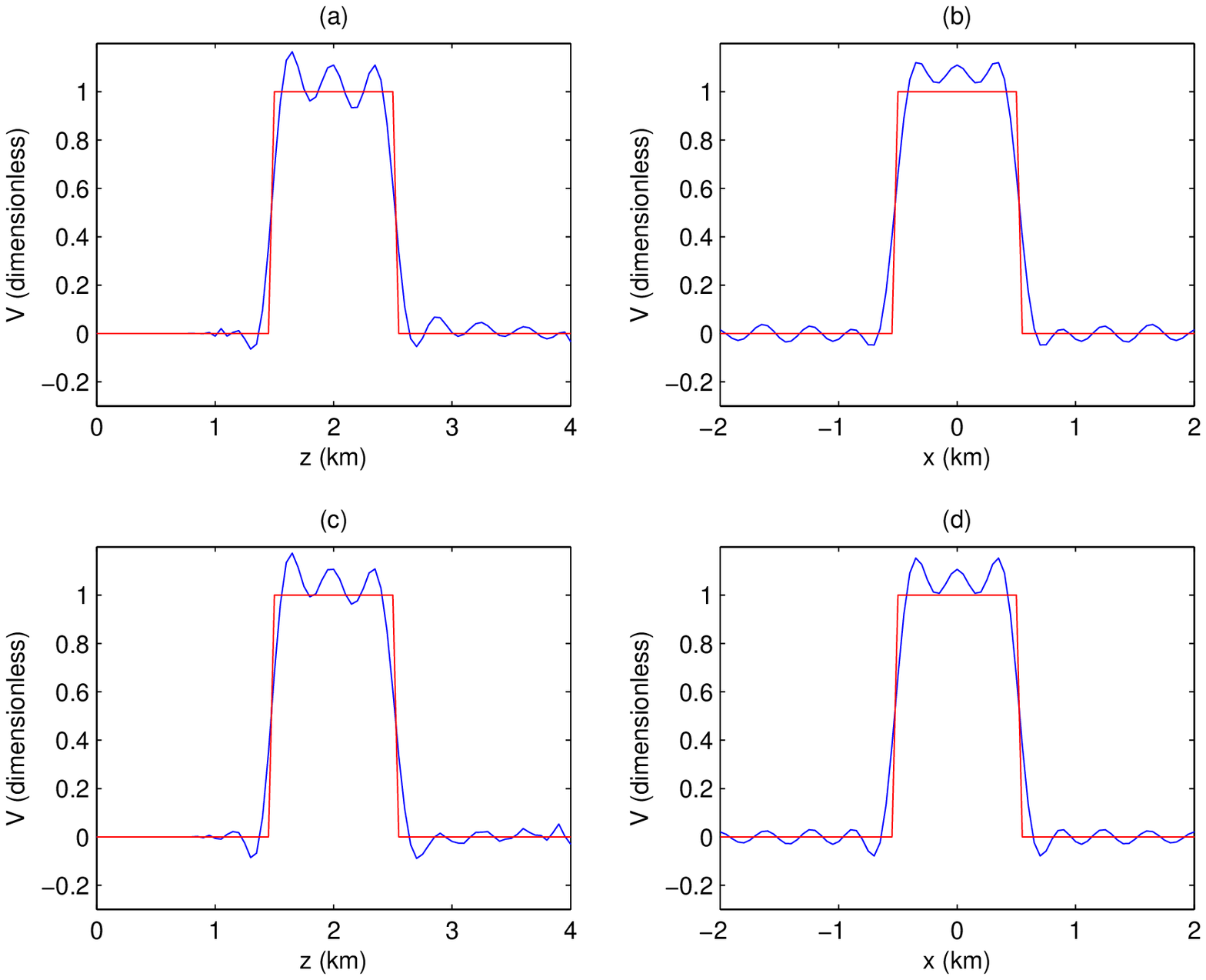,width=\wid,clip=}
\end{center}
\caption{Cross-sections through the results of Fig. 5.  (a,b) Vertical and horizontal slices through the unrestricted result.
(c,d) Vertical and horizontal slices through the restricted results.}
\end{figure}

\begin{figure}[!htbp]
\begin{center}
\epsfig{file=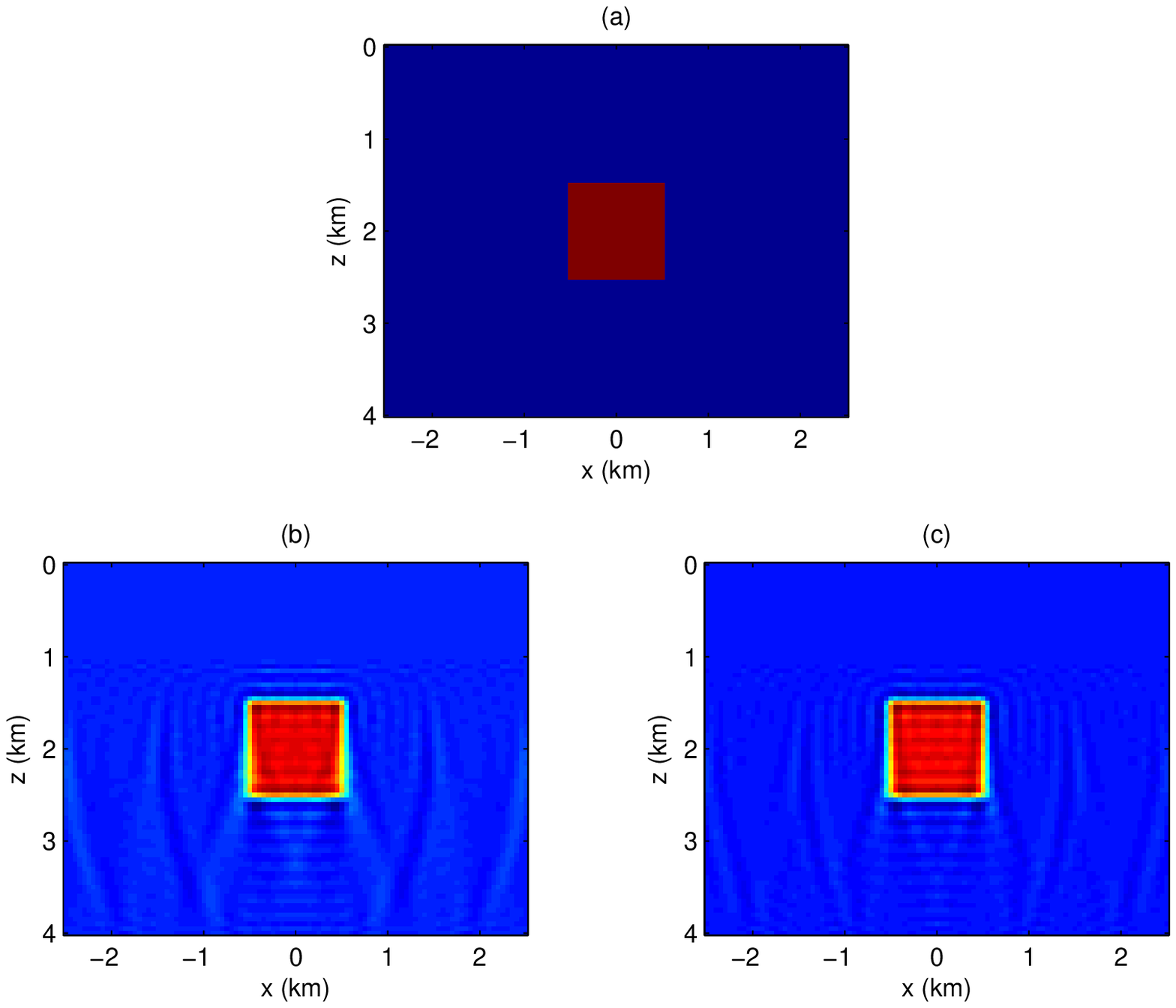,width=\wid,clip=}
\end{center}
\caption{Results for the two-layer background model but with frequencies 1,2,...,10 Hz used.  (a) True model of $V$.  
(b,c) Results for unrestricted and restricted cases.}
\begin{center}
\epsfig{file=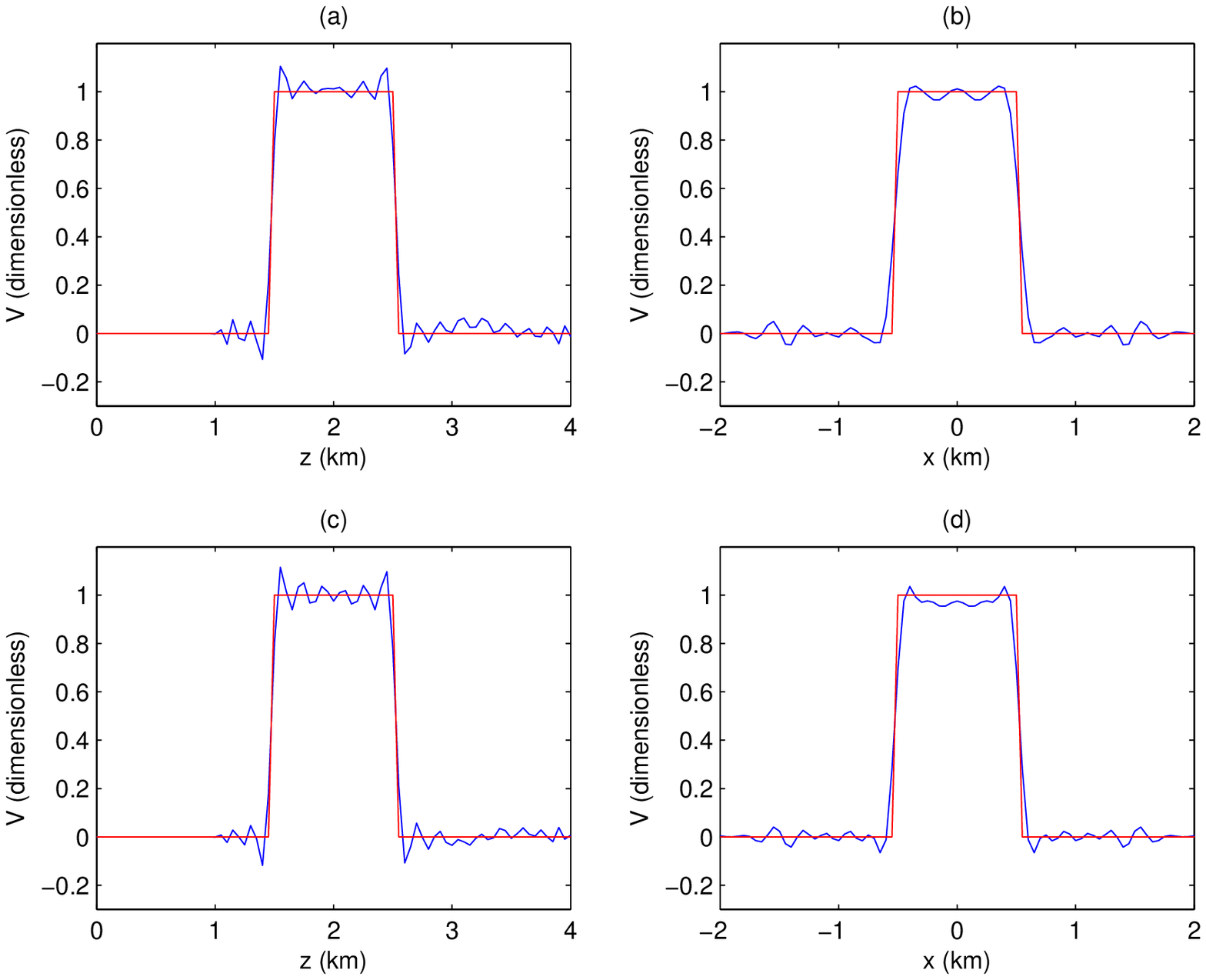,width=\wid,clip=}
\end{center}
\caption{Cross-sections through the results of Fig. 7.  (a,b) Vertical and horizontal slices through the unrestricted result.
(c,d) Vertical and horizontal slices through the restricted results.}
\end{figure}

\begin{figure}[!htbp]
\begin{center}
\epsfig{file=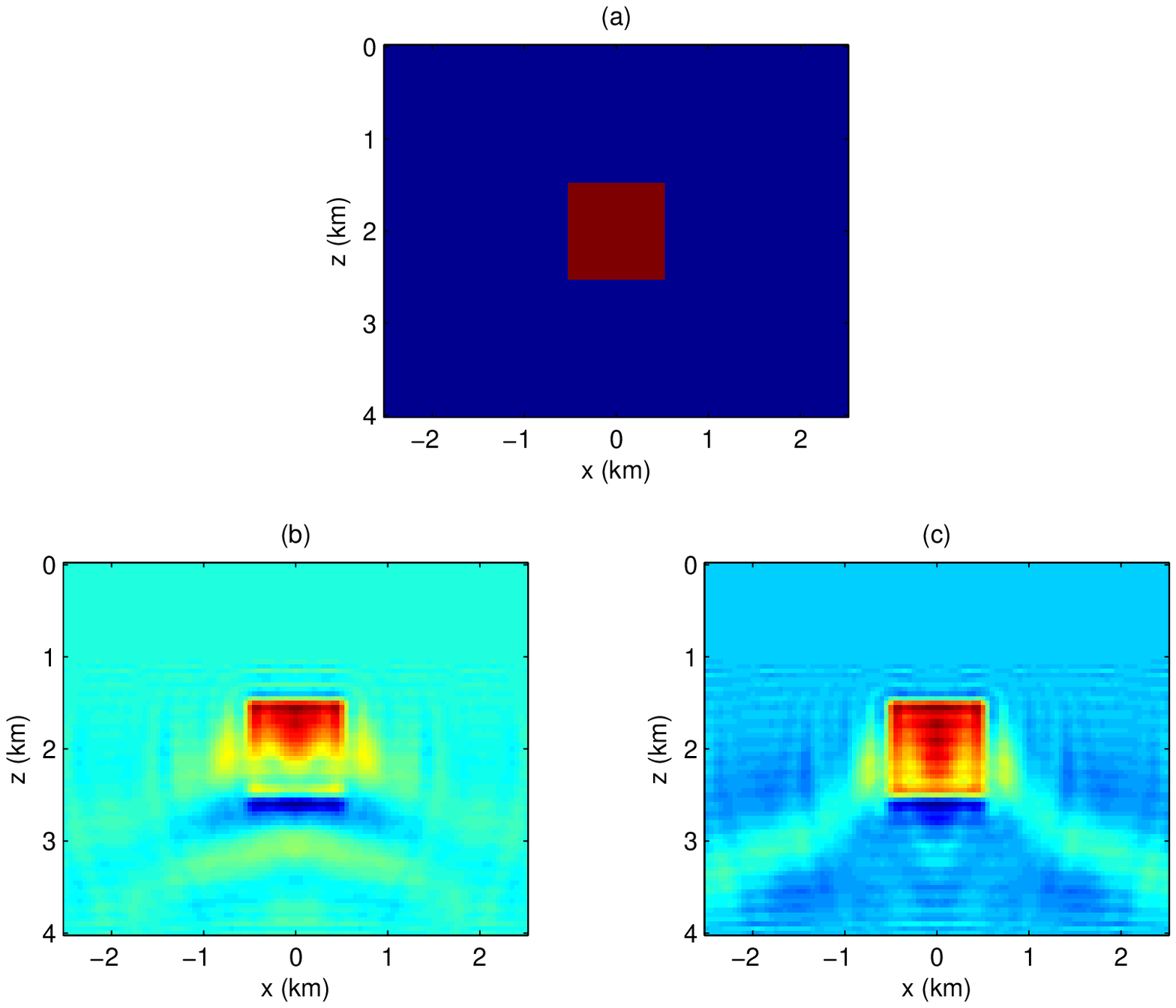,width=\wid,clip=}
\end{center}
\caption{Results for the two-layer background model but with frequencies 5,6,...,10 Hz used.  (a) True model of $V$.  
(b,c) Results for unrestricted and restricted cases.}
\begin{center}
\epsfig{file=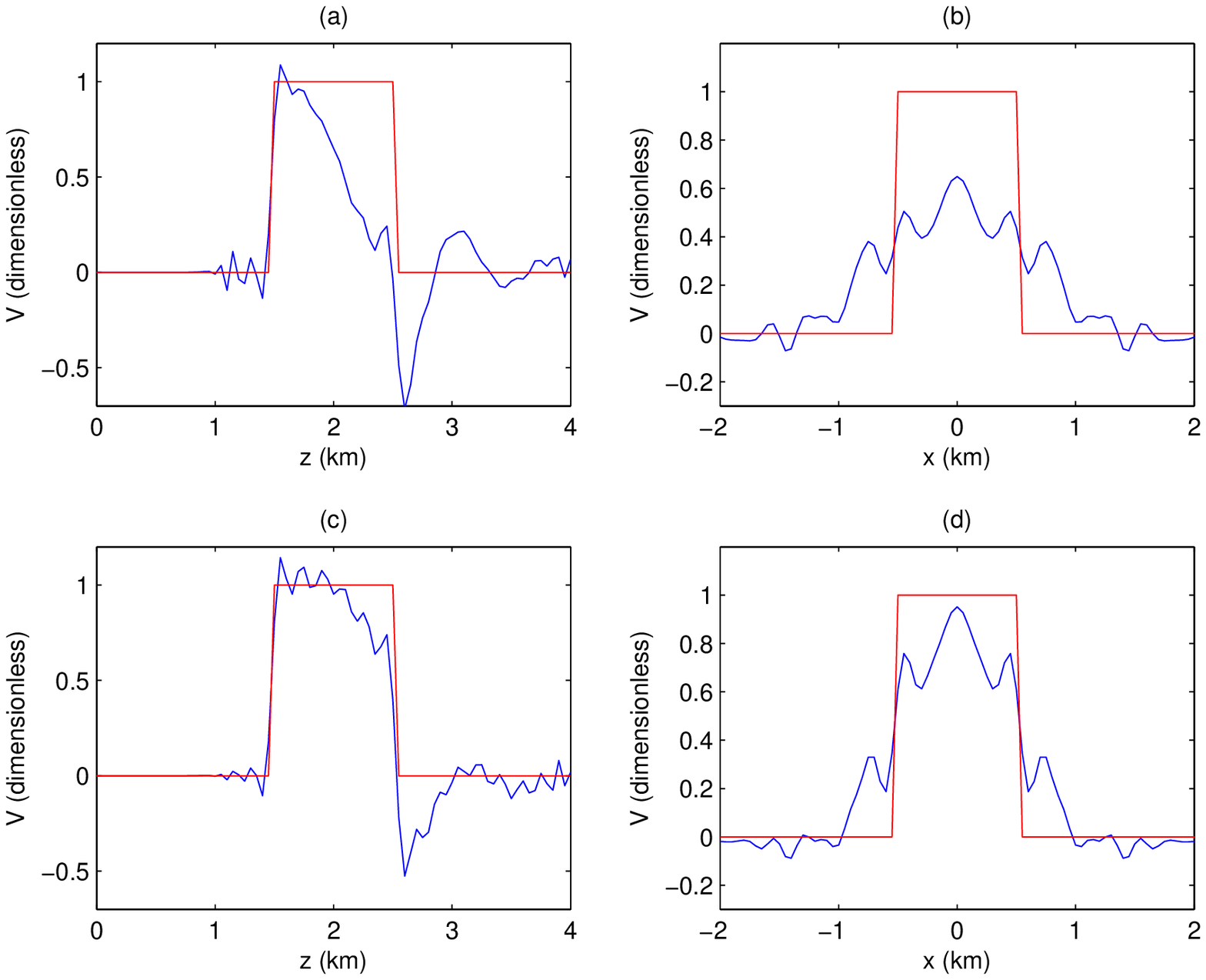,width=\wid,clip=}
\end{center}
\caption{Cross-sections through the results of Fig. 9.  (a,b) Vertical and horizontal slices through the unrestricted result.
(c,d) Vertical and horizontal slices through the restricted results.}
\end{figure}

\begin{figure}[!htbp]
\begin{center}
\epsfig{file=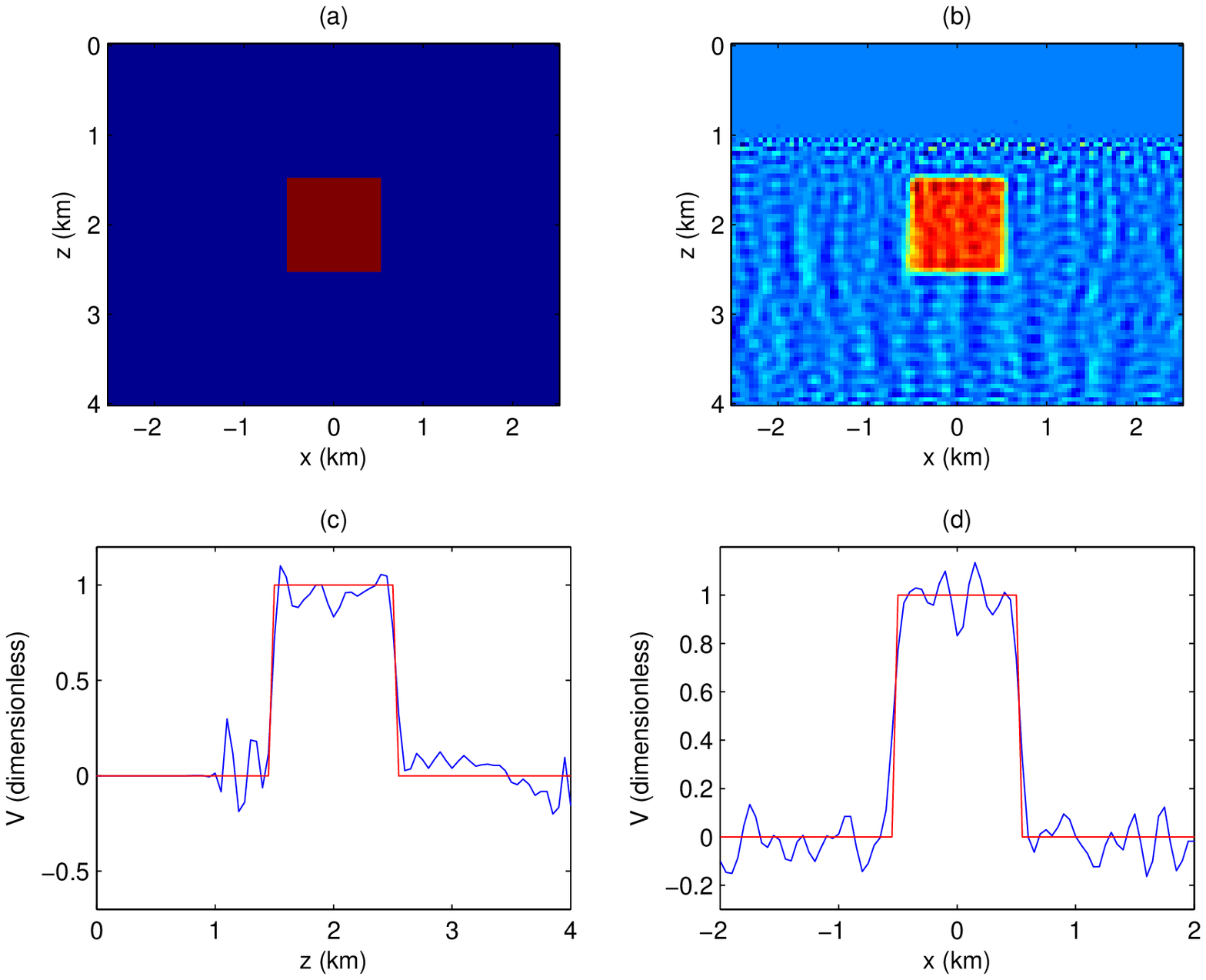,width=\wid,clip=}
\end{center}
\caption{Results for the two-layer background model for 1,2,...,10 Hz and 5\% noise added. (a) True model of $V$. (b) Reconstruction
result. (c,d) Vertical and horizontal slices through the result.}
\end{figure}

In real seismic exploration data, we are usually limited to frequencies larger than 5 Hz.  In Figures 9 and 10, we see the results
of extended diffraction tomography when we use only 5-10 Hz frequencies and the same two-layer model as before.  We do see degradation
of the results, but $\kz$ restriction helps to subdue them partially.  We hypothesize that using an even more realistic 1D reference
model, one with a gradient in the lower half-space, would significantly improve the results because of the presence of turning
rays, which would sample yet more effectively the components of the spectrum of $V$ that are presently most difficult to obtain.  Future
work will explore this hypothesis.

Our last example demonstrates the robustness of our method.  Here we test the effect of adding 5\% noise.  By that,
we mean that for each frequency, we added to $D(m,h;\w)$ complex-valued numbers that had a constant absolute value of .05 times the 
maximum absolute value of $D$ at that frequency but had a phase that was uniformly randomly distributed between 0 and $2\pi$.

In Figure 11, we see the effect of adding the noise to our data for the two-layer reference model.  In this case, we used the full 
1-10 Hz frequency range. Although there is now noise in the recovered image, our use of a large tolerance in the SVD inversion has 
preserved the vast majority of the structure.

%%%%%%%%%%%%%%%%%%%%%%%%%%%%%%%%%%%%%%%%%%%%%%%%%%%%%%%%%%%%%%%%%%%%%%%%%%%%%%%%%%%%%%%%%%%%%%%%%%%%%%%%%
\section{Discussion}

We have presented the development of extended diffraction tomography, a new variant of diffraction tomography that can very efficiently
and often very accurately solve the linear seismic inversion problem for surface reflection data with finite aperture.  The algorithm, 
when properly implemented, exhibits excellent robustness with respect to random noise.  In addition, we have extended the theory
to a realistic 3D acquisition geometry.

The main strength of extended diffraction geometry originates from the use of laterally invariant (i.e., 1D) reference models.  This allows
the decomposition of the full 2D inverse problem into many independent 1D problems, making the method naturally highly parallelizable. 
In contrast to classical diffraction tomography, the use of 1D reference models also allows for greater realism and accuracy in the
inversion.  We have already demonstrated that a simple two-layer reference model greatly enhances the results.  Further work 
will be done to confirm that even more realistic models containing gradients will yield yet more substantial improvements.

The reduction of the inverse problem to the solution of relatively small matrix equations opens the possibility of applying the
formidable tools of numerical linear algebra.  We have already demonstrated that the simple restriction of the allowed values of the
vertical wavenumber $\kz$ yields significant benefits.  Future efforts may include the incorporation of more sophisticated methods
such as Tikhonov and total variation regularizations (see e.g. Aster {\em et al.}, 2005).

Further testing of the method must also be done to explore how it handles data that is {\em not} generated via first-order Born
scattering, i.e., we must see how it handles multiples, in order to evaluate its usefulness in nonlinear inversion.  Although this 
method may have some value in and of itself, its ultimate purpose is to be used in the full-waveform nonlinear inversion of realistic
seismic data.

%%%%%%%%%%%%%%%%%%%%%%%%%%%%%%%%%%%%%%%%%%%%%%%%%%%%%%%%%%%%%%%%%%%%%%%%%%%%%%%%%%%%%%%%%%%%%%%%%%%%%%%%%
\section*{Acknowledgments}

I wish to thank Ru-Shan Wu for many interesting and stimulating discussions.  I also thank the sponsors of the WTOPI consortium for 
their support of this work.

%%%%%%%%%%%%%%%%%%%%%%%%%%%%%%%%%%%%%%%%%%%%%%%%%%%%%%%%%%%%%%%%%%%%%%%%%%%%%%%%%%%%%%%%%%%%%%%%%%%%%%%%%


\begin{thebibliography}{}

\bibitem{1} Aster, R. C., B. Borchers, and C. H. Thurber, 2005, Parameter Estimation and Inverse Problems: Elsevier
Academic Press.

\bibitem{2} Devaney, A. J., 1984, Geophysical diffraction tomography: Trans., Inst. Electr, Electron. Eng., GE-22, 3--13.

\bibitem{3} Devaney, A. J., and M. L. Oristaglio, 1984, Geophysical diffraction tomography: SEG Expanded Abstracts,
{\bf 3}, 330--333.

\bibitem{4} Pratt, R. G., 1999a, Seismic waveform inversion in the frequency domain, Part 1: Theory and verification
in a physical scale model: Geophysics, {\bf 64}, 888--901.

\bibitem{5} Pratt, R. G., 1999b, Seismic waveform inversion in the frequency domain, Part 2: Fault delineation in
sediments using crosshole data: Geophysics, {\bf 64}, 901--913.

\bibitem{6} Press, W. H., S. A. Teukolsky, W. T. Vetterling, and B. P. Flannery, 1992, Numerical Recipes in Fortran: Cambridge
University Press.

\bibitem{7} Schlottmann, R. B., 2006, Direct waveform inversion via iterative inverse propagation: SEG Expanded Abstracts, 
{\bf 25}, 2146--2150.

\bibitem{8} Sirgue, L., and R. G. Pratt, 2004, Efficient waveform inversion and imaging: A strategy for
selecting temporal frequencies: Geophysics, {\bf 69}, 231--248.

\bibitem{9} Weglein, A. B., F. V. Araujo, P. M. Carvalho, R. H. Stolt, K. H. Matson, R. Coates, D. Corrigan, D. J. Foster, 
S. A. Shaw, and H. Zhang, 2003, Inverse scattering series and seismic exploration: Inverse Problems, {\bf 19}, R27--R83.

\bibitem{10} Wu, R.-S., and M. N. Toks\H{o}z, 1987, Diffraction tomography and multi-source holography applied to seismic
imaging: Geophysics, {\bf 52}, 11--25.

\end{thebibliography}
\end{document}